\documentclass[runningheads]{llncs}
\usepackage{graphicx}
\usepackage{amsmath}
\begin{document}
\title{DNA-GCN: Graph convolutional networks for predicting DNA-protein binding}

\titlerunning{DNA-GCN}

\author{Yuhang Guo\inst{1,}\thanks{Equal Contribution.} \and
Xiao Luo \inst{1,2,*} \and
Liang Chen\inst{1} \and
% Tao Li\inst{1} \and
Minghua Deng\inst{1}}
\authorrunning{Guo et al.}
\institute{School of Mathematical Sciences, Peking University, Beijing, China \and
Damo Academy, Alibaba Group, Hangzhou, China
\\
\email{\{yuhangguo, xiaoluo, clandzyy, dengmh\}@pku.edu.cn }}
\maketitle              % typeset the header of the contribution
\begin{abstract}

Predicting DNA-protein binding is an important and classic problem in bioinformatics. Convolutional neural networks have outperformed conventional methods in modeling the sequence specificity of DNA-protein binding. However, none of the studies has utilized graph convolutional networks for motif inference. In this work, we propose to use graph convolutional networks for motif inference. We build a sequence k-mer graph for the whole dataset based on k-mer co-occurrence and k-mer sequence relationship and then learn DNA Graph Convolutional Network(DNA-GCN) for the whole dataset. Our DNA-GCN is initialized with a one-hot representation for all nodes, and it then jointly learns the embeddings for both k-mers and sequences, as supervised by the known labels of sequences. We evaluate our model on 50 datasets from ENCODE. DNA-GCN shows its competitive performance compared with the baseline model. Besides, we analyze our model and design several different architectures to help fit different datasets. 

\end{abstract}
\section{Introduction}

DNA-binding proteins play an important role in gene regulation. It's well-known that the transcription of each gene is controlled by a regulatory region of DNA relatively near the transcription start site. There are two fundamental components in transcription, the short DNA regulatory element, and its corresponding gene regulatory proteins. DNA binding sites are small but highly variable, which makes them difficult to detect. Several experimental methods were developed(e.g. ChIP-seq\cite{zhang2008model}) to solve this problem, but they are usually costly, and each has its artifacts, biases, and limitation. Based on sequence-based data, the problem of predicting DNA-protein binding is to model the sequence specificity of protein binding (i.e. connect a relationship between sequence-based data and binary labels of data). Specifically, the task is a classification problem given training  DNA sequences and their binary labels to predict labels of given testing sequences in the dataset. Recent work on motif inference includes conventional machine learning-based methods(e.g. SVM, Random Forest)\cite{lee2016ls,corrado2016rnacommender,ghandi2014enhanced} and deep learning-based methods(e.g. CNN, RNN)\cite{alipanahi2015predicting,zeng2016convolutional,quang2016danq,shen2018recurrent}. CNN's and RNNs have shown their superiority compared with conventional machine learning-based methods. However, when it comes to small datasets, the performance of the models is often limited. Besides, the models of CNN's usually learn truncated motifs that aren't desired\cite{blum2019neural}.

On the other hand, the binary labels of the sequences are up to whether they have some specific regions called a motif. If we regard the "A", "C", "G" and"T" as special kinds of characters, k-mer can be treated as words and DNA sequences can be viewed as sentences. The k-mer related to given motifs can be viewed as keywords and predicting DNA-protein binding is transformed into the problem of text classification. 

In this paper, we propose a novel method based on Graph Convolutional Networks\cite{kipf2016semi} -- DNA-GCN for predicting DNA-protein binding. In DNA-GCN, firstly we construct a single large graph from the whole dataset, and then GCN is utilized to obtain neighborhood information. By this, predicting DNA-protein binding is turned into a semi-supervised node classification problem. We choose a lot of different datasets with limited samples to evaluate our model. The model shows competitive performance on the task of predicting TF binding sites. All code is public in \url{https://github.com/Tinard/dnagcn}. In summary, our contributions in this paper are twofold. (1) We propose a novel graph convolutional network for predicting DNA-protein binding. To the best of our knowledge, we are the first to model the sequence specificity with a graphical model and utilize GCN to learn the sequences and k-mer embeddings. (2) The empirical results show that our proposed model has a competitive performance compared with the baseline models on many datasets with limited sequences. We suppose that our method could contribute to the study of DNA sequence modeling and other biological models.

\section{Related Work}

\subsection{Deep learning for motif inference}

The deep learning method for motif inference can be categorized into two groups -- CNN-based and RNN-based methods. As for the CNN-based model(may contain RNN), DeepBind\cite{alipanahi2015predicting} is the first CNN-based model to predict DNA-protein binding and since then deep learning has been widely utilized in this field for its great performance. \cite{zeng2016convolutional} shows that deploying more convolutional kernels is always beneficial. iDeepA\cite{pan2017attention} applies an attention mechanism to automatically search for important positions. DeeperBind\cite{hassanzadeh2016deeperbind} and iDeepS\cite{pan2018prediction} add an LSTM layer on DeepBind to learn long dependency within sequences to further improve the prediction performance. As for the RNN-based model, KEGRU\cite{shen2018recurrent} identifies TF binding sites by combining Bidirectional Gated Recurrent Unit (GRU) network with k-mer embedding. Besides model selection, CONCISE\cite{avsec2017modeling} and iDeep\cite{pan2017rna} integrate other information(e.g. structured information) into predicting RBP-binding sites and preference. Other work includes data augmentation\cite{cao2018simple}, circular filters\cite{blum2019neural} and convolutional kernel networks\cite{chen2019biological,luo2019deepprune,luo2020expectation}.

\begin{figure}
    \centering
    \includegraphics[width=\textwidth]{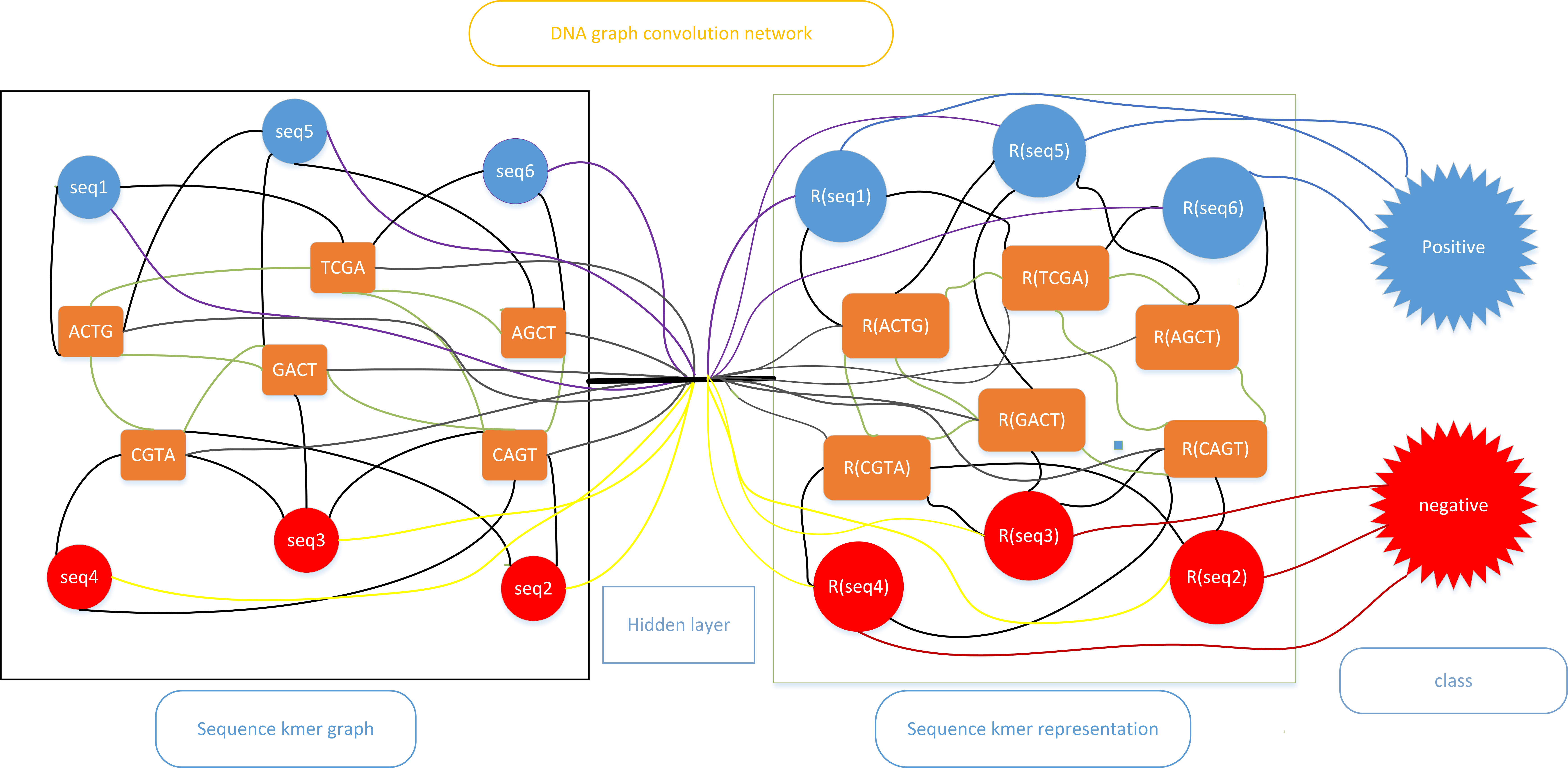}
    \caption{Architecture of DNA-GCN.} 
    \label{fig:1}
\end{figure}

\subsection{Graph Convolutional Networks}

In the past few years, graph convolutional network\cite{kipf2016semi} has attracted wide attention for its learning hierarchical information on graphs\cite{li2018deeper,chen2018fastgcn,hamilton2017inductive}. It has been shown that the GCN model is a special form of Laplacian smoothing which makes the features of nodes similar to their neighbors and makes the subsequent classification task much easier. The layer-wise propagation rule is formulated as:
$$
H ^ { ( k + 1 ) } = \phi \left( \tilde { D } ^ { - \frac { 1 } { 2 } } \tilde { A } \tilde { D } ^ { - \frac { 1 } { 2 } } H ^ { ( k ) } W ^ { ( k ) } \right)
$$
where $H^(k)$ is the node representation matrix, $W^(k)$ is the trainable parameter matrix of the $k$-th layer, $H^(0)=X$ is the origin feature matrix, $\tilde { A} = A + I $ is the adjacency matrix with increased self-connection, $ \ tilde {D} $ is the degree matrix of $\tilde{A}$ (i.e. $\tilde { D } _ { i i } = \sum _ { j } \tilde { A } _ { i j }$), and $ \ phi $ is the activation function, such as $ RELU (\ cdot) $. In addition, there are many variants of GCN, focusing on improving the performance of GCN and coping with the storage bottleneck of GCN\cite{chen2018fastgcn,hamilton2017inductive,velivckovic2017graph,wu2019simplifying,zhuang2018dual}. In recent years, GCN has been used to handle many tasks, such as text classification, drug recommendation and laboratory test classification \cite{yao2018graph, mao2019medgcn,li2018classifying}, and has shown better performance than baseline on different tasks.

\subsection{Heterogeneous graph}

The heterogeneity is an intrinsic property of a heterogeneous graph, i.e., various types of nodes and edges. Apparently, different types of nodes have different features which fall in different feature space. However, if we utilize Laplacian smoothing directly, different feature space is mixed which seems unreasonable. HAN\cite{wang2019heterogeneous} utilizes an attention mechanism to generate node embedding by aggregating features from meta-path-based neighbors in a hierarchical manner. MedGCN\cite{mao2019medgcn} assumes that there is no edge between nodes in the same type and the propagation rule is rewritten as
$$H _ { i } ^ { ( k + 1 ) } = \phi \left( \sum _ { j = 1 } ^ { n } A _ { i j } \cdot H _ { j } ^ { ( k ) } \cdot W _ { j } ^ { ( k ) } \right)$$ 
We assume that the number of types is $n$, $A_{ij}$ is the adjacency matrix between nodes type $i$ and $j$. $W^{(k)}_j$ is the learnable weight matrix for type $i$ nodes in layer $k$.

\section{DNA-GCN}

\subsection{Sequence k-mer graph}

First of all, we construct a large and heterogenous graph $G = ( \mathcal { V } , \mathcal { E } )$ where $\mathcal { V } , \mathcal { E } $ are sets of nodes and edges respectively to describe the relationship between sequences and k-mers.  As shown in Figure \ref{fig:1}, nodes in the graph have two types: sequences nodes and k-mers nodes. The number of nodes in sequence k-mer graph $|\mathcal { V }|$ is the number of sequences including training set, validation set, and testing set plus the number of possible k-mers. The weight of the edge between sequences and k-mers is the number of occurrences of k-mer multiplied by inverse sequence frequency(ISF) in the sequence. ISF is the logarithmically scaled inverse fraction of the number of sequences that contain the k-mer. We test two models with or without ISF and found that the former is better. We suppose that some common k-mers which isn't related to the motif in real-world data may disrupt the performance and ISF can ease the effect of irrelevant k-mers. Point-wise mutual information(PMI) is utilized to calculate weights between two k-mers. Above all, the adjacent matrix of sequence k-mer graph is formulated as:
$$
A _ { i j } = \left\{ \begin{array} { l l } { \operatorname { PMI } ( i , j ) } & { i , j \text { are k-mers } , \operatorname { PMI } ( i , j ) > 0 } \\ { \mathrm { O } * \operatorname { ISF } _ { i j } } & { i \text { is sequence, } j \text { is k-mer } }  \\ { 0 } & { \text { otherwise } } \end{array} \right.
$$
We can computed $PMI(i,j)$ as 
$$
\begin{aligned} \operatorname { PMI } ( i , j ) & = \log \frac { p ( i , j ) } { p ( i ) p ( j ) } \\ p ( i , j ) & = \frac { \# W ( i , j ) } { \# W } \\ p ( i ) & = \frac { \# W ( i ) } { \# W } \end{aligned}
$$
where $\#W(i)$ is the number of sequences that contain k-mer $i$, $\#W (i, j)$ is the number of sequences that contain both k-mer $i$ and $j$, and $\#W$ is the total number of sequences in the dataset. We set $k$ to be $4$, and because a motif length is between 6 and 20, the information of a motif may be spitted into several k-mers. We believe that if two k-mers co-occur in a sequence frequently, they are probably to co-decide whether a sequence contains a motif. From the formulation above, a positive PMI value indicates a high correlation of k-mers while a negative PMI value indicates little or no correlation. From the analysis, we set positive PMI values to be the weights of edges between k-mers.

\subsection{DNA-GCN}

After constructing the sequence k-mer graph, we feed the graph into GCN. Because the information of sequences is embedded into the graph, we set the feature matrix $X =I$ for simplicity. At first, we feed the graph into a two-layer GCN, and the second layer node embeddings only have one dimension and then are fed into a softmax classifier for classification.
$$
Z = \operatorname { softmax } \left( \hat { A } \operatorname { ReLU } \left( \hat { A } X W _ { 0 } \right) W _ { 1 } \right)
$$
where $\hat{A}=\tilde { D } ^ { - \frac { 1 } { 2 } } \tilde { A } \tilde { D } ^ { - \frac { 1 } { 2 } }$ and $softmax(x_i)= \frac { 1 } { \mathcal { Z } } \exp \left( x _ { i } \right)$ with partition function $\mathcal { Z } $(i.e. $\mathcal { Z } = \sum _ { i } \exp \left( x _ { i } \right)$). 
The model has transformed into a semi-supervised model this time and the cross-entropy error over all labeled sequences determine loss function. 
$$\mathcal { L } = - \sum _ { d \in \mathcal { Y } _ { D } }  Y _ { d } \ln Z _ { d  }$$
where $\mathcal { Y } _ { D }$ is the set of sequence indices that have labels and $Y$ is the label indicator vector.

We give an ideal example. For the dataset with a specific protein, its positive samples contain the motif "AACGTC" while negative samples don't contain it. AACG, ACGT, CGTC are the key 4-mers which guide classification, the three k-mers is connected to all the positive sample. By training guided by labeled sample, the three k-mers can be trained to have the features that point to positive label, and then the model can predict the label of a testing sequence by whether the key k-mers is connected to target sequences (i.e. Features of key k-mers can be transferred to testing sequences or not). Overall, the information is transferred from labeled samples to k-mers, and then from k-mers to unlabeled sequences. From the analysis above, we need at least two times of Laplacian smoothing(i.e. two layers GCN) to construct our DNA-GCN. By experiments and experience that too many layers lead to over-smoothing features, we set the number of layers to be two. 

We also notice that the sequence k-mer graph is a heterogeneous graph. As a result, feeding our graph into GCN directly seems to be unreasonable. The layer-wise propagation rule can be rewritten as
$$
H ^ { ( k + 1 ) } = \phi \left( \hat{A}(H ^ { ( k ) } W ^ { ( k )) } \right).
$$
The formulation above means we feed every node into an identical single-layer perceptron, and from this point of view, we can feed each type of node into a specific perception. We assume that $n_1$ nodes represent sequences, $n_2$ nodes represent k-mers and $n=n_1+n_2$. $\tilde{A}$ can be partitioned into two blocks, $(\tilde{A_1})_{n_1\times n}$ and $(\tilde{A_2})_{n_2\times n}$ according to node type. $W_1^k$ and $W_2^k$ are weight matrices with the same shape. The layer-wise propagation rule for a heterogeneous graph is formulated as
$$
H ^ { ( k + 1 ) } = \phi \left( \tilde{A}_1(H ^ { ( k ) } W _1^ { ( k )})+\tilde{A}_2(H ^ { ( k ) } W _2^ { ( k )})  \right).
$$
We call the second method DNA-HGCN. It's evident that in DNA-HGCN, the number of parameters double. As a result, DNA-HGCN is easier to overfit. We also utilize Simple Graph Convolutional Network(SGC)\cite{wu2019simplifying} to build our model. SGC removes nonlinearities and collapses weight matrices between consecutive layers to reduce excess complexity while keeping the model's performance.

\subsection{Implementation of DNA-GCN}

Our model is trained using Adam\cite{kingma2014adam} optimizer with a learning rate of 0.001 for 10000 epochs. $20\%$ sequences are chosen from the labeled set to construct the validation set. We chose the best model according to the performance in the validation set. We set the embedding size of the first convolution layer as 100.

We utilized the area under the ROC(AUC)\cite{davis2006relationship,fawcett2004roc} to assess the prediction performance of prediction. Our model is implemented using Tensorflow\cite{abadi2016tensorflow} for Python.

\section{Result}

\subsection{Datasets}

To evaluate the performance of our model, we chose the 50 ChIP-seq ENCODE datasets. Each of these datasets corresponds to a specific DNA-binding protein (e.g., transcription factor); its positive samples are 101bp DNA sequences which were experimentally confirmed to bind to this protein, and its negative samples were created by shuffling these positive samples. All these datasets were downloaded from \url{http://cnn.csail.mit.edu/}. We didn't test our performance in all available datasets for two reasons. On one hand, there is a tendency that when training sets contain more samples, better performance can be obtained by DeepBind. From this view, raising performance on these large datasets with high performance (i.e. AUC is about $99\%$) is useless. On the other hand, our model needs to be learned with the presence of both training and test data. Moreover, the recursive neighborhood expansion across layers poses time and memory challenges for training with large graphs(i.e. large datasets). As a result, We selected 50 datasets with limited samples to test the performance of our model.  

\begin{figure}
    \centering
    \includegraphics[width=10cm]{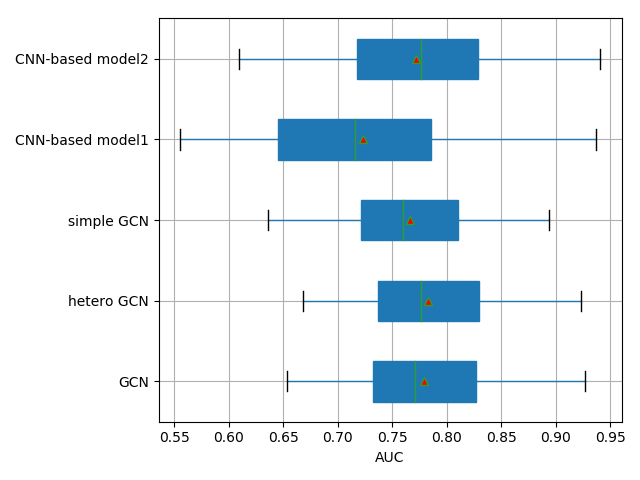}
    \caption{Overall performance of DNA-GCN.}
    \label{fig:2}
\end{figure}
\subsection{Baselines}

\paragraph{Gkm-SVM}\cite{ghandi2014enhanced} introduces alternative feature sets using gapped k-mers and develops an efficient tree data structure for computing the kernel matrix. Compared to original kmer-SVM and alternative approaches, gkm-SVM predicts functional genomic regulatory elements and tissue-specific enhancers with significantly improved accuracy. 

\paragraph{CNN-based model}\cite{zeng2016convolutional} is similar to the architecture of DeepBind. We utilize their best model with 128 convolutional kernels as our baseline. The overall performance of the CNN-based model is better than DeepBind in ENCODE datasets. Model 1 and model 2 refer to CNN-based model with 1 and 128 convolutional kernels, respectively.

\subsection{Our model outperforms on many datasets}

\begin{figure}
    \centering
    \includegraphics[width=10cm]{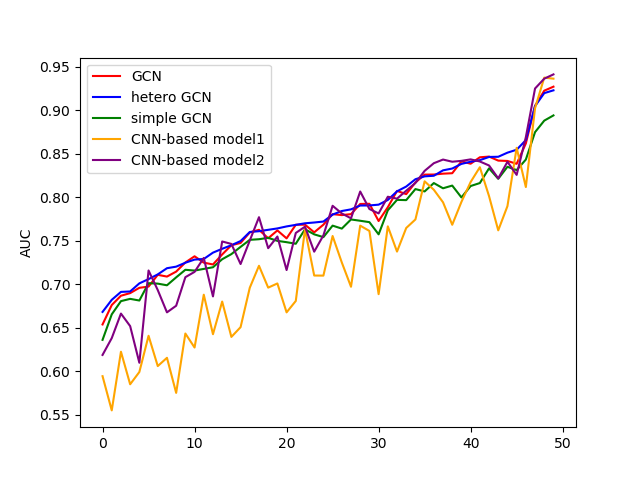}
    \caption{Performance of DNA-GCN on specific datasets.}
    \label{fig:3}
\end{figure}

From Figure \ref{fig:2}, the performance of DNA-HGCN among the datasets is sightly better than the best baseline. DNA-GCN-based three different models have close performance. DNA-GCN based on SGC sacrifices a little performance for its least calculation while DNA-HGCN has better performance with doubled parameters. 

As for specific datasets[Figure \ref{fig:3}], we found that the performance among the CNN-based model and GCN-based model is inconsistent, which shows that GCN and CNN predict DNA-protein binding from different views. The performance of the same kind of models is consistent, regardless of the GCN-based model or CNN-based model. We believe if two methods can be combined, the best performance can be arrived at.

\section{Conclusion}

In this paper, a novel method named DNA-GCN is proposed to predict DNA-protein binding. We build a heterogeneous sequence k-mer graph for the whole dataset and turn predicting the DNA-protein binding problem into a node classification problem. Our DNA-GCN can transit information from labeled sequences to key k-mers and predict labels of unlabeled sequences. By experiments, we show that a simple two-layer GCN brings up promising results by comparing numerous models on many datasets with a limited number of sequences. 

One the other hand, we believe that we haven't made the best use of GCN to predict DNA-protein binding. Much improvement may be achieved by adjusting the architectures and hyper-parameters on a given dataset. Although DNA-GCN can't arrive at motif logos like CNN-based model, it can provide us with information about which k-mers are important in the classification. Above all, our model gives a brand new perspective to study the motif inference. 

\section*{Acknowledgements}
This work was supported by The National Key Research and Development Program of China (No.2016YFA0502303) and the National Natural Science Foundation of China (No.31871342).

%
% the environments 'definition', 'lemma', 'proposition', 'corollary',
% 'remark', and 'example' are defined in the LLNCS documentclass as well.
%


{\small
\bibliographystyle{ieee}
\bibliography{ms}
}

\end{document}